\documentclass[a4paper,10pt]{spie}

\usepackage{graphicx}
\usepackage{noitemsep}

\usepackage{verbatim}
\usepackage{float}

\usepackage[latin1]{inputenc}
\usepackage[T1]{fontenc}
\usepackage{ae}
\usepackage{amsmath, amssymb}
\usepackage{color}
\usepackage{parskip}
\usepackage{tabularx}
\usepackage{xspace}
\usepackage[german, english]{babel}

\newcommand{\sa}{superattenuator}

\newcommand{\FP}{Fabry-Perot}
\newcommand{\por}{power recycling}

\newcommand{\prc}{power-re\-cycling cavity}

\newcommand{\pom}{power-recycling mirror}

\newcommand{\Mi}{Michelson interferometer}

\newcommand{\mc}{mode cleaner}
\newcommand{\mHc}{mode-cleaner}
\newcommand{\omc}{output mode cleaner}

\newcommand{\gw}{gravitational wave}

\newcommand{\gHw}{gravi\-ta\-tion\-al-wave}

\newcommand{\mSec}[1]{Section~\ref{#1}}

\newcommand{\M}[1]{\ensuremath{{\rm TEM}_{#1}}}

\newcommand{\pd}{photo diode}

\newcommand{\mFig}[1]{Figure~\ref{#1}}

\newcommand{\ie}{\mbox{i.\,e.}\xspace}

\let \IG \includegraphics

\title{Status of VIRGO*}
\author{F.~Acernese\supit{6}, 
P.~Amico\supit{10}, 
N.~Arnaud\supit{8},
S.~Avino\supit{6},
D.~Babusci\supit{4}, 
R.~Barill\'e\supit{2}, 
F.~Barone\supit{6}, 
L.~Barsotti\supit{11}, 
M.~Barsuglia\supit{8},
F.~Beauville\supit{1}
% ...
\skiplinehalf
\supit{1}Laboratoire d'Annecy-le-Vieux de Physique des Particules, Annecy-le-Vieux, France;\\
\supit{2}European Gravitational Observatory (EGO), Cascina (Pi), Italia;\\
\supit{3}INFN, Sezione di Firenze/Urbino, Sesto Fiorentino, and Universit\`a di Firenze, and Osservatorio Astrofisico di Arcetri, Firenze and Universit\`a di Urbino, Italia;\\
\supit{4}INFN, Laboratori Nazionali di Frascati, Frascati (Rm), Italia;\\
\supit{5}SMA, IPNL, Villeurbanne, Lyon, France;\\
\supit{6}INFN, sezione di Napoli and Universit\`a di Napoli "Federico II" Complesso Universitario di Monte S.Angelo, and Universit\`a di Salerno, Fisciano (Sa), Italia;\\
\supit{7}Observatoire de la C\^ote d'Azur, D\'epartement Fresnel Interf\'erom\'etrie Laser pour la Gravitation et l'Astrophysique, Nice, France;\\
\supit{8}Laboratoire de l'Acc\'el\'erateur Lin\'eaire (LAL), IN2P3/CNRS-Univ.   de Paris-Sud, Orsay, France;\\
\supit{9}ESPCI, Paris, France;\\
\supit{10}INFN, Sezione di Perugia and Universit\`a di Perugia, Perugia, Italia;\\
\supit{11}INFN, Sezione di Pisa and Universit\`a di Pisa, Pisa, Italia;\\
\supit{12}INFN, Sezione di Roma and Universit\`a "La Sapienza",  Roma, Italia.}

\authorinfo{Send correspondence to A. Freise, E-mail: andreas.freise@ego-gw.it}

\begin{document}
\maketitle

\vspace{.2cm}
$^*$ Presented by A. Freise for the VIRGO Collaboration.

\begin{abstract}
The French-Italian interferometric gravitational wave
detector VIRGO is currently being commissioned. Its 
principal instrument is a Michelson interferometer with 3 km
long optical cavities in the arms and a power-recycling mirror. This
paper gives an overview of the present status of the system.
We report on the presently attained sensitivity and the system's performance 
during the recent commissioning runs. 

After a sequence of intermediate stages, the interferometer 
is now being used in the so-called recombined configuration.
The input laser beam is spatially filtered by a 144 m long 
input mode-cleaner before being injected to the main interferometer. 
The main optics are suspended from so-called \sa s, which 
provide an excellent seismic isolation. The two 3 km long Fabry-Perot 
arm cavities are kept in resonance with the laser light, and the
Michelson interferometer is held on the dark fringe. An automatic mirror 
alignment system based on the Anderson technique has been implemented 
for the arm cavities. The light leaving the dark port contains the 
gravitational wave signal; this light is filtered by an output 
mode-cleaner before being detected by a photo detector.
This setup is the last step on the way to the final 
configuration, which will include power recycling.
\end{abstract}


\section{INTRODUCTION}
The French-Italian collaboration VIRGO~\cite{VIRGO:prop} 
has built a large-scale interferometric \gHw\ detector
near Pisa, Italy. The main instrument is a Michelson interferometer (MI) with 3\,km long
\FP\ cavities in its arms. A passing gravitational wave can be detected as a change
in the relative length between a set of free falling test masses. 
High-quality optics are suspended to act as quasi-free
test masses at the end of the \Mi\ arms so that a \gw, passing
perpendicular to the detector, will be detected in the 
interferometer signal. The \FP\ cavities in the arms enhance the light
power and thus increase the optical gain of the interferometer.
\begin{figure}[h]
\begin{center}
\IG [scale=.6, angle=0] {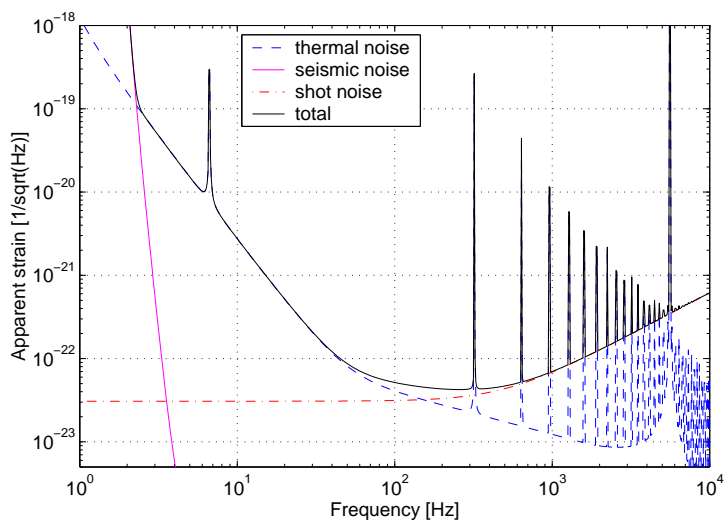}
\end{center}
\caption{\label{fig:sens}The expected sensitivity of the VIRGO detector: The spectral densities 
of the limiting noise sources are shown as an apparent strain. The total gives the sensitivity, limited by
seismic disturbances below 3\,Hz, by thermal noise up to 100\,Hz and by shot noise
for higher frequencies.}
\end{figure}

In order to reach the extreme sensitivity required 
for detecting \gw s (see \mFig{fig:sens}), the VIRGO detector uses special techniques to minimise the
coupling of noise into the interferometric signal. The large optics (mirrors
and beam splitters) are super-polished
fused silica pieces with very low absorption and scattering. They are located
in an ultra-high vacuum system and suspended from a sophisticated seismic isolation system.
The laser light is stabilised in frequency and spatially filtered
by a 144\,m long suspended \mc.

The end of the construction phase was marked by the installation of the last
test mass in June 2003.
Between 2002 and 2003, the \emph{central interferometer} (CITF)~\cite{CITF} was commissioned,
and in 2003 the commissioning of the VIRGO detector was finally started.
Step by step the
various subsystems are being implemented and tested. 
In fact, then the two 3\,km long arm cavities were put
into operation and then combined to the large-scale \Mi. 
To date the system is run in the
so-called recombined configuration, which is the last milestone before the detector
will reach its final configuration including \por.

The first scientific data-taking periods are planned for 2005. At that point VIRGO will
join LIGO, TAMA and GEO\,600~\cite{ligo,tama,geo}
in the existing network of interferometric detectors.

In this paper, the status of the detector will be presented with special emphasis on its
performance during the recent data-taking periods.

\section{Optical Layout}
\begin{figure}[h]
\begin{center}
\IG [viewport=0 0 450 400,clip,scale=.8, angle=0] {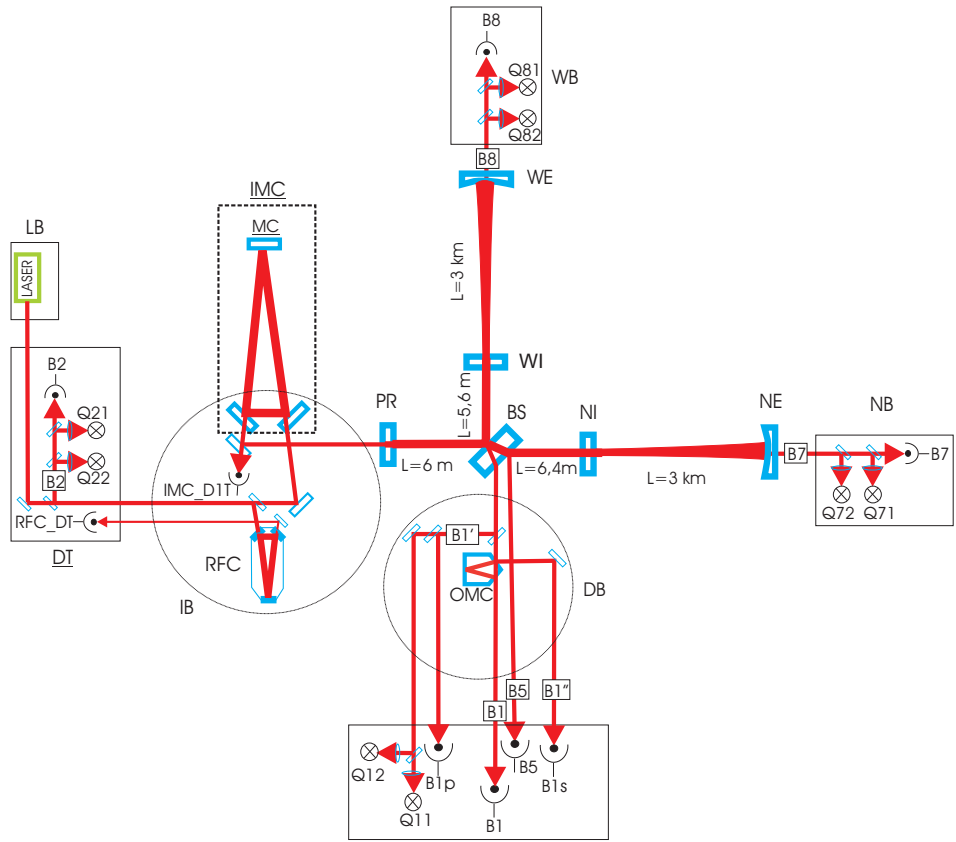}
\end{center}
\caption{\label{fig:optical-layout}A simplified schematic of the
optical design for VIRGO: The laser beam is directed on the \emph{detectors table}
(DT) into the first vacuum chamber, the injection tower, in which all
optical components are attached to a suspended optical bench, the \emph{injection
bench} (IB). After passing the \emph{input \mc} (IMC), the beam is injected
through the \emph{\pom} (PR) into the main interferometer. The beam is split and
enters the two 3\,km long arm cavities, the \emph{west arm} (WA) and 
\emph{north arm} (NA). The \Mi\ (MI) is held at the dark fringe so that
most of the light power is reflected back to the \emph{\pom} (PR). Together with the MI the PR forms a 
\FP-like cavity in which
the light power is enhanced. The light from the
dark port of the beam splitter is filtered by an \emph{\omc} (OMC) before being detected on a set
of 16 photo diodes (B1), which generate the main output signal of the detector. 
The other photo diodes shown in this schematic with names starting with {\bf B}
are used for longitudinal control of the interferometer; diodes named with a {\bf
Q} represent split photo detectors used for alignment control.}
\end{figure}

\mFig{fig:optical-layout} shows a simplified optical layout of the VIRGO interferometer
in its final configuration.
The laser light, 20\,W @1064nm provided by an injection-locked master-slave
solid state laser (Nd:YAG), enters the vacuum system at the 
\emph{injection bench} (IB). The beam is spatially filtered by a 144\,m long input \mHc\ cavity (IMC)
before being injected into the main interferometer. 

The laser frequency is pre-stabilised using the IMC cavity as a reference. The
low-frequency stability is achieved by an additional control system that stabilises
the IMC length below 15\,Hz to the length of a so-called reference 
cavity (RFC). The RFC is a 30\,cm long rigid triangular cavity suspended in vacuum; it 
consists of ULE, a material with a very low thermal expansion coefficient.

A beam with 10\,W of power enters the Michelson interferometer through the \pom. 
It is split
into two beams that are injected into the 3\,km long arm cavities. Together with the MI 
(which is held on the dark fringe) the 
\pom\ forms a \FP-like cavity in which the light power is resonantly enhanced.
The optical (recycling) gain of this cavity is 50, which leads to 500\,W of light at the beam splitter; 
the bandwidth (FWHM) of the \prc\ (PRC) is designed to be $\approx 20$\,Hz.

The finesse of the arm cavities is approximately 50; this yields a circulating light power
of 8\,kW. The MI is held on the dark fringe, and the \gw\
signal is expected in the beam from the dark port, which is leaving the vacuum
via the so-called detection bench (DB). The detection bench is a suspended optical
bench accommodating several optical components. The main beam is passed through an
\emph{\omc} (OMC), a 2.5\,cm long rigid cavity. 

To achieve a shot noise limited sensitivity, the main output
beam is detected by a group of 16 InGaAs 
\pd s. For control purposes, other \pd s are also used. Useful signals are obtained
by detecting the light in the transmission of the
arm cavities and in reflection of the \prc, as well as detecting the reflection at the 
secondary surface of the beam splitter and the light reflected by the OMC.

\section{Commissioning of the VIRGO detector}
The commissioning period is divided into three main phases with different
detector setups: in phase A the two arm cavities 
are used separately, whereas in phase B the arm cavities and the beam splitter
are controlled to hold the Michelson on the dark fringe; this is the so-called
\emph{recombined} configuration. In both phases the \pom\
is kept misaligned. In phase C the \prc\
will also be aligned to reach the final VIRGO
configuration. According to plan, the commissioning is to be finished by
the end of this year with the detector in its final configuration.
\begin{figure}[h]
\begin{center}
\IG [scale=0.5, angle=0] {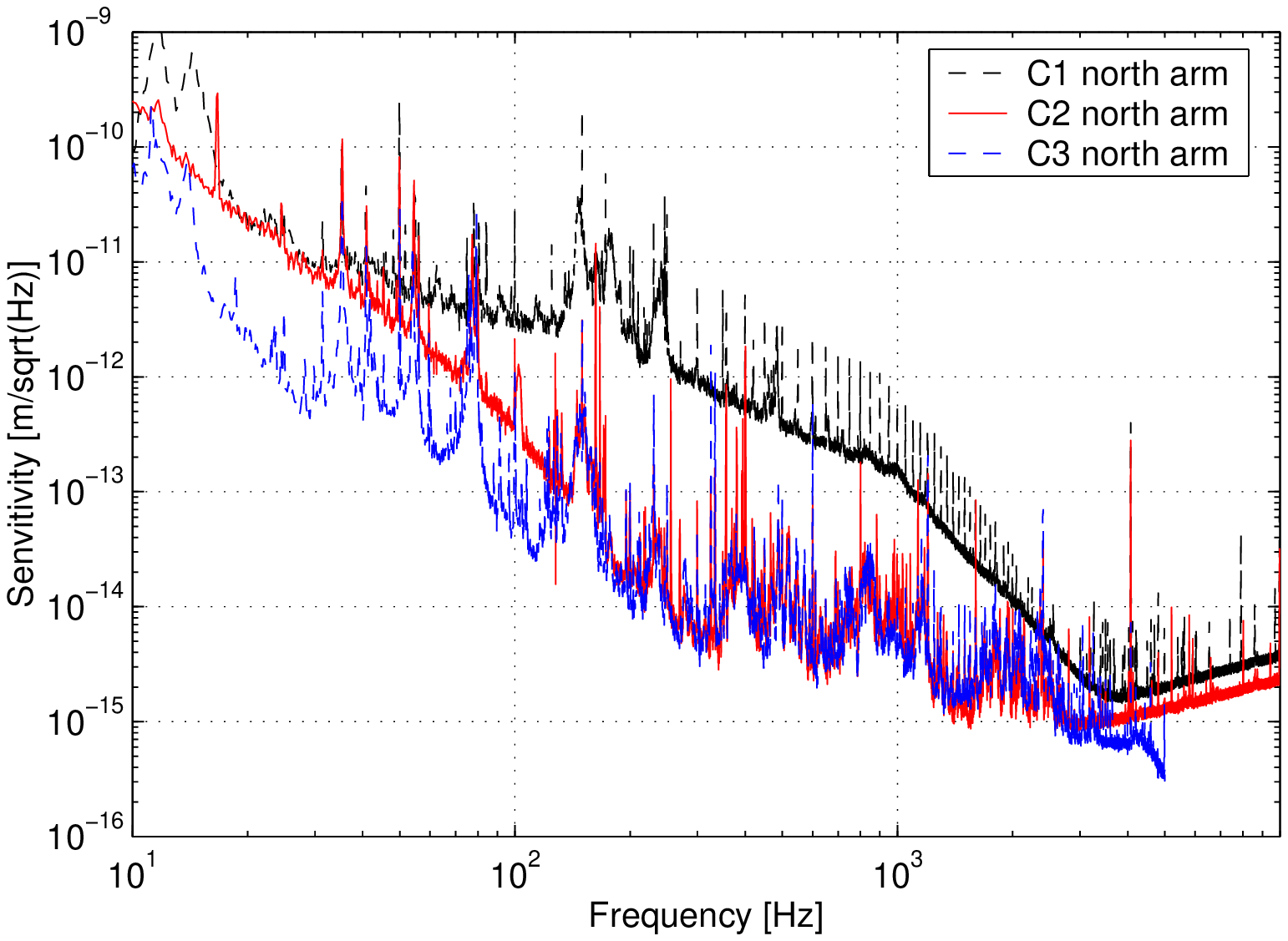}\hspace{7mm}\IG [scale=0.5, angle=0] {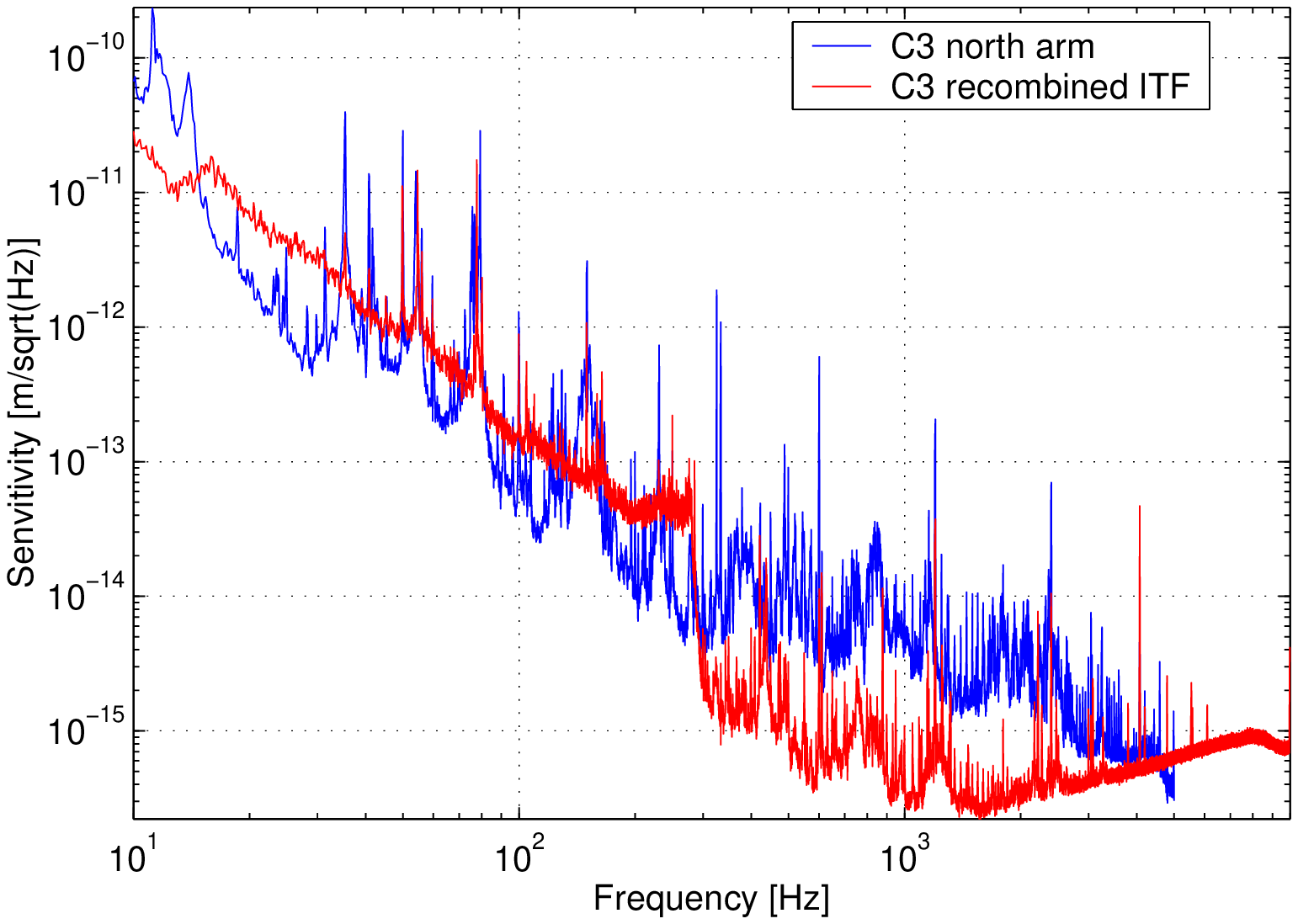}
\end{center}
\caption{\label{fig:Csens}These graphs show three plots of the displacement
sensitivity of the detector during the three last commissioning runs.
Currently the sensitivity is limited by technical noise, \ie laser
frequency noise and electronic (sensor) noise. This is expected 
for the intermediate optical configurations.
The three traces in the left plot show the sensitivity of the north arm locked only.
This configuration is basically measuring the frequency noise which is 
limiting the sensitivity from a few Hz to the kHz region. The 
improvement in sensitivity in this configuration could be achieved by 
optimising the IMC control loops. The right plot shows a comparison
between the sensitivity of the north arm and the 
sensitivity for the recombined interferometer in which the arm cavities
and the \Mi\ are controlled (see text).}
\end{figure}

The go-ahead for the commissioning of the detector was given by
the alignment of the IMC output beam through the 3\,km long arms in September 2003.
In October the north arm cavity was locked, i.e. the cavity length was 
actively controlled to be resonant with the input laser frequency. A
stable operation of the control was achieved at the first trial using a
lock acquisition algorithm and loop filter that were previously
tested with a SIESTA\footnote{SIESTA is a time-domain simulation developed
in the VIRGO collaboration, able to model the optical and mechanical
properties of the interferometer as well as control loops.} simulation.

The west arm cavity became available (i.e. the optics were aligned
and locally controlled) in November 2003, and the first longitudinal lock
was achieved in late December. In the meantime, the automatic alignment of the
mirrors of the north arm had been implemented, so that in January
2004, the automatic alignment control could be started for the first time
on one of the long cavities.

In February the first lock of the recombined interferometer could
be demonstrated. It turned out that some of the control signals
were polluted by offsets that depend on the alignment of the main
optics. Consequently the work on the recombined interferometer 
was stopped in order to implement
the automatic alignment of the west arm cavity, which was
finished in April 2004. At the same time the second stage
of frequency stabilisation (see \mSec{sec:freq}) had been implemented, and its first
successful operation was demonstrated in March 2004.

We have now completed phase B: the
arm cavities can be controlled independently in all
degrees of freedom, and the recombined configuration
has been successfully tested and is currently used
to characterise the optical setup. 
Currently the light power of the IMC output beam is 8\,W. 
The detector is used without recycling, \ie the \pom\
is misaligned by several mrads and merely attenuates the transmitted light ($T_{\rm PR}=8\%$). 
This yields a light power at the beam splitter of $\approx 0.6$\,W and 
approximately 10\,W of light circulating in the
arm cavities.

Approximately every two months a short period of continuous data taking, a 
so-called \emph{commissioning run} is scheduled.
The data show the progress of the commissioning of the instrument and
the effects of the detector evolution on the detector sensitivity.

The following
runs have been performed so far:
\begin{itemize}
\item{C1: 14.--17. November 2003, north arm cavity longitudinally controlled}
\item{C2: 20.--23. February 2004, north arm cavity with longitudinal and angular
control}
\item{C3: 23.--27. April 2004, two configurations: a) north arm as in C2 plus
a second stage of laser frequency stabilisation, and b) the recombined 
interferometer (no automatic alignment nor second stage of frequency stabilisation)}
\end{itemize}
\begin{figure}[t]
\begin{center}
\IG [scale=.5, angle=0] {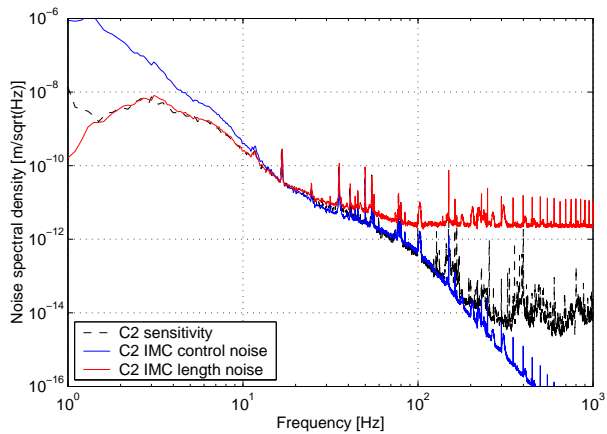}
\end{center}
\caption{\label{fig:noise}Noise investigations for data taken during the C2 run: 
This graph shows the projection of 
IMC length and control noise 
into the gravitational wave channel. (Note that the projection models used
for this plot are valid only in a selected frequency band: 
the IMC length noise projection is valid only for Fourier frequencies $<10$\,Hz and the 
IMC control noise projection for frequencies $>10$\,Hz.)
This shows that during the C2 run the frequency noise limits
the sensitivity of the detector between 2\,Hz and 200\,Hz.}
\end{figure}
\mFig{fig:Csens} shows the displacement sensitivity of the detector during
these three runs. Compared to the final detector the intermediate
optical setups are more sensitive to frequency noise, so that this noise
limits the current sensitivity over a wide frequency range. In addition, the 
electronic noise (or sensor noise) is larger than the shot noise because of the
low circulating light power without recycling.
The C1 sensitivity has been found to be limited by laser frequency noise
from 7\,Hz to 4\,kHz and by electronic noise for higher frequencies.
Therefore, the main goal between C1 and C2 was to reduce 
the laser frequency noise in the measurement band ($>10$\,Hz).
By reducing
the bandwidth of IMC control loops (local control, automatic 
alignment, length control) the frequency noise could be
reduced in the range $>10$\,Hz by slightly enhancing it
for lower frequencies. One can see the result in the improved sensitivity
in C2, which is still mainly limited by laser frequency noise
(from 2\,Hz to 200\,Hz) as explained in \mFig{fig:noise}.
For frequencies from 200\,Hz to a few kHz the noise could
not be clearly identified yet. Its origins are 
partly seismic noise near the injection system and 
mechanic vibrations by vacuum pumps. Very likely this noise couples into the
main output also through the laser frequency.
Two different configurations were used during C3. At first only
the north arm was controlled, but with a second stage of frequency
stabilisation. For a single cavity this frequency stabilisation
yields in principle no sensitivity improvement; the slight improvement as seen
in \mFig{fig:Csens} is simply due to a lower bandwidth of the
RFC loop (see below for description of the frequency stabilisation
control).

It is interesting to compare the two sensitivity curves referring to the
C3 run. The recombined interferometer has an intrinsically higher rejection
of frequency noise. This can be seen in the better sensitivity at high frequencies.
At frequencies below 300\,Hz, electronic noise is limiting
the sensitivity. This noise is re-injected by the arm cavity controls, which,
in this run, used preliminary sensor signals.

In the following, we will 
give an overview of the main subsystems, 
with emphasis on the improvements achieved so far during the commissioning.

\section{Control and data acquisition}
The control of the main optical components is mainly performed by two
digital control systems: for each suspension a digital signal processor
with a sampling of 16 bits at 20\,kHz
(DSP) is used to read the local sensors (for example, accelerometers, LVDTs, optical
lever signals) and generate the correction signals for the local control
and the inertial damping. In addition, the so-called \emph{Global Control} (GC)~\cite{GC} 
is used to read photo diode signals and generate global correction signals (for
example the differential motion of the cavities). The correction signals are 
sent to the DSP of the respective suspensions via a digital optical link (DOL).
The DSP finally sends them to the appropriate actuator.

The data acquisition chain was successfully tested during the commissioning runs.
The current data rate is 7\,Mbytes/s of compressed data.  

\section{Suspensions}
While the VIRGO detector uses an optical layout similar to that of other interferometric detectors, it aims 
at achieving very high sensitivities down to low frequencies (>10\,Hz). This is possible due
to a sophisticated suspension system for the main optical components, the 
\sa\ (SA)~\cite{SA}. This suspension system resides in up to 9\,m tall vacuum chambers and
is composed of several filter stages. The mirror and a reference mass 
each are suspended by two steel wires
from a so-called marionette. Longitudinal forces to the mirror can be applied
via coil-magnet actuators, with the magnets attached to the mirror surface and the
coils being supported by the reference mass. In addition, coil-magnet actuators at the
marionette level can be used to apply forces in longitudinal and angular directions. 


The topmost filter is rigidly connected 
to a ring that rests on three legs forming an inverted pendulum. This pendulum
has a very low horizontal resonance frequency (40\,mHz). While providing a 
good attenuation of seismic noise already at low frequencies, it also allows 
to move the top point of the suspension by up to
$\pm$20\,mm using very small forces.

At the resonance frequencies of the various isolation stages, the seismic motion
is actually enhanced. Therefore, active controls are used to damp the mirror motion
at these frequencies (DC to approximately 5\,Hz). The control is split into two
parts: the inertial damping~\cite{ID} (ID) and the local control~\cite{LC} (LC).
The ID is a control system that uses three linear 
variable differential transformers (LVDTs) and three accelerometers located 
on the upper mechanical filter. The actuation is performed via three coil-magnet
actuators also located on the same mechanical filter.

The LC can control the mirror in three degrees of freedom: the displacement along the optical
axis ($z$), the angular rotation around the vertical axis ($\Theta_y$), and the
rotation around the horizontal axis perpendicular to the beam ($\Theta_x$). 
The feedback for angular control is applied only via the coil-magnet actuators on the marionette,
whereas longitudinal feedback can be sent to
the marionette, to the reference mass actuators and to the
top stage of the inverted pendulum (see below).


The excellent passive seismic isolation of the SA was characterised during
the commissioning of the CITF~\cite{CITF}. Measurement of the seismic attenuation
yielded a factor of $10^{-8}$ at 4\,Hz as an upper limit. The RMS motion of the mirror
was measured to be below 1\,$\mu$m and 1\,$\mu$rad for longitudinal and angular
displacement respectively. 

\begin{figure}[h]
\begin{center}
\IG [viewport= 0 0 1100 370,clip,scale=0.36, angle=0] {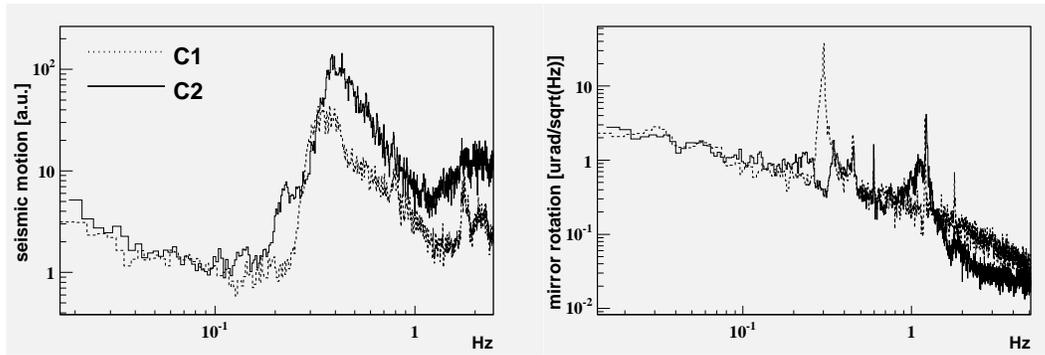}
\end{center}
\caption{\label{fig:seismic}These plots demonstrate the improvement
of the ID control between C1 to C2. The left graph shows a spectrum of the horizontal 
seismic motion
at some time during the respective run. At both times a storm at the
site was giving rise to a large seismic noise, especially around
300\,mHz. The right plot shows the angular motion ($\Theta_y$) of the 
suspension at the mirror level for the same time. It can be seen
that during C1 the larger seismic disturbance caused excess mirror motion
and that after the upgrade of the control in C2 the storm 
caused no increase in the angular motion of the mirror.}
\end{figure}

During the commissioning the control loops of the ID and LC were
optimised for better performance. \mFig{fig:seismic} shows a comparison of
two data stretches recorded during the C1 and C2 runs: during both runs
the seismic noise was enhanced due to a storm, especially around 300--500\,mHz, 
by a factor of approximately 50.
The C1 data show that this noise is visible as a large motion 
of the mirrors at 300\,mHz, associated with an angular resonance of the payload. 
Consequently the filters of the control were adjusted
and in C2 the large seismic noise could successfully be damped. 

The earth tides cause a slow regular elongation of the
arm cavities of $\approx 200\cdot10^{-6}$\,m amplitude (peak-to-peak).
The dynamic range of the coil-magnet actuators at the
mirror level (used for longitudinal feedback)
is just $\approx 100\cdot10^{-6}$\,m. To achieve continuous 
stable operation one has to increase the dynamic range
of the control. During the CITF a hierarchical feedback
system was tested, which sent low-frequency feedback (bandwidth
70\,mHz) to the
top stage of the suspension. It has been shown that this
not only provides a compensation for the earth tides but
also reduces the RMS of the correction signals by one order
of magnitude~\cite{CITF-top}.

During the last commissioning runs this \emph{tidal control} was not
available so that a regular manual adjustment of the top stage
was performed by operators. Recently a first implementation of the
tidal control was demonstrated on one arm and will be used in the
future as standard configuration.

\section{Frequency Stabilisation}\label{sec:freq}
The extreme sensitivity needed for the detection of gravitational waves requires
a sophisticated laser stabilisation. For example, the typical frequency noise of the
principal laser source is given with 1\,kHz$/\sqrt{\rm Hz}$ at 10\,Hz. In order to reach the desired
sensitivity of VIRGO this frequency noise has to be reduced to $3\cdot10^{-5}$\,Hz$/\sqrt{\rm Hz}$
(at the beam splitter).
Such large suppression can only be achieved by an active stabilisation.
\begin{figure}[h]
\begin{center}
\IG [scale=.23, angle=0] {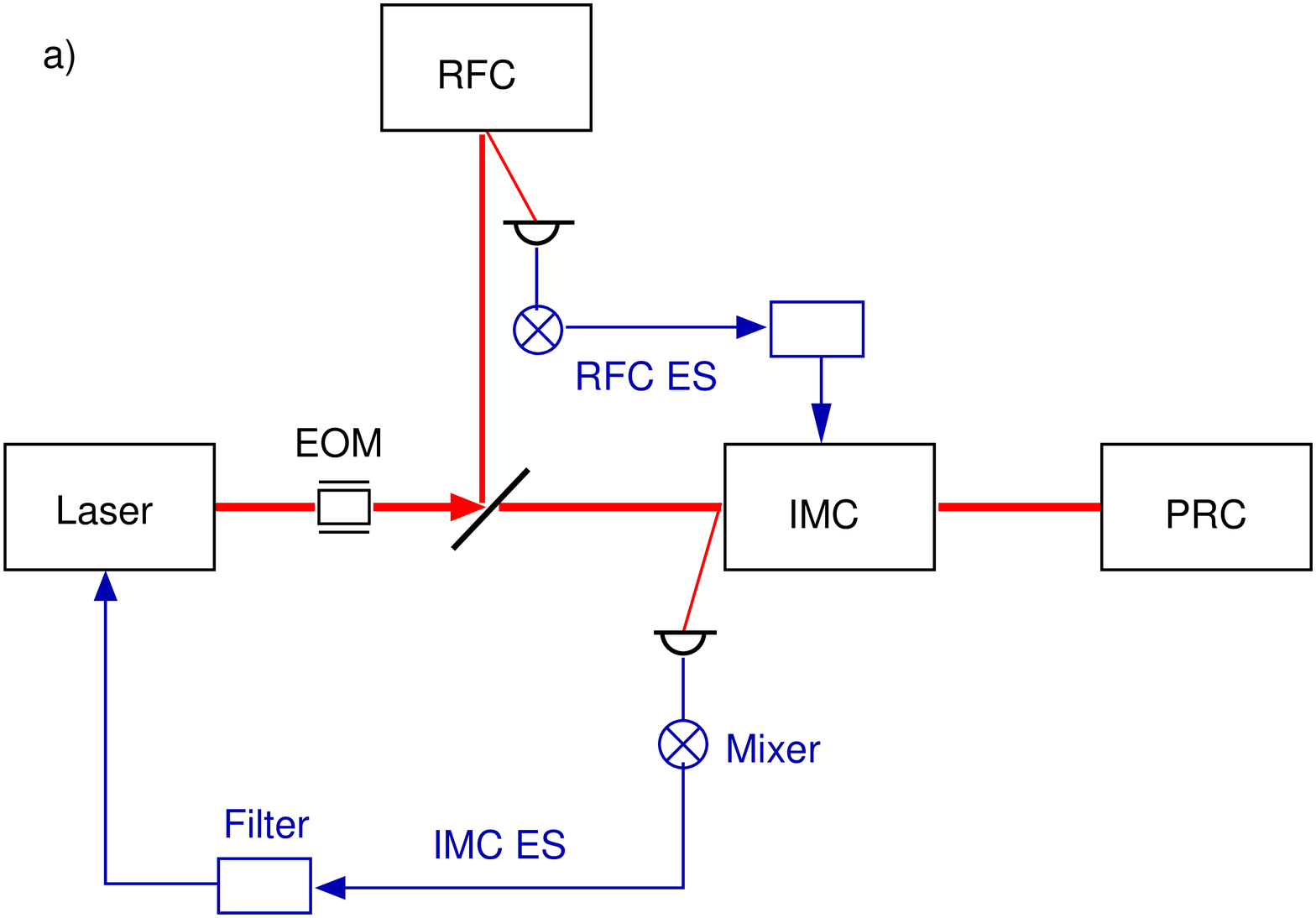}\hspace{5mm}\vrule\hspace{5mm}\IG [scale=.23, angle=0] {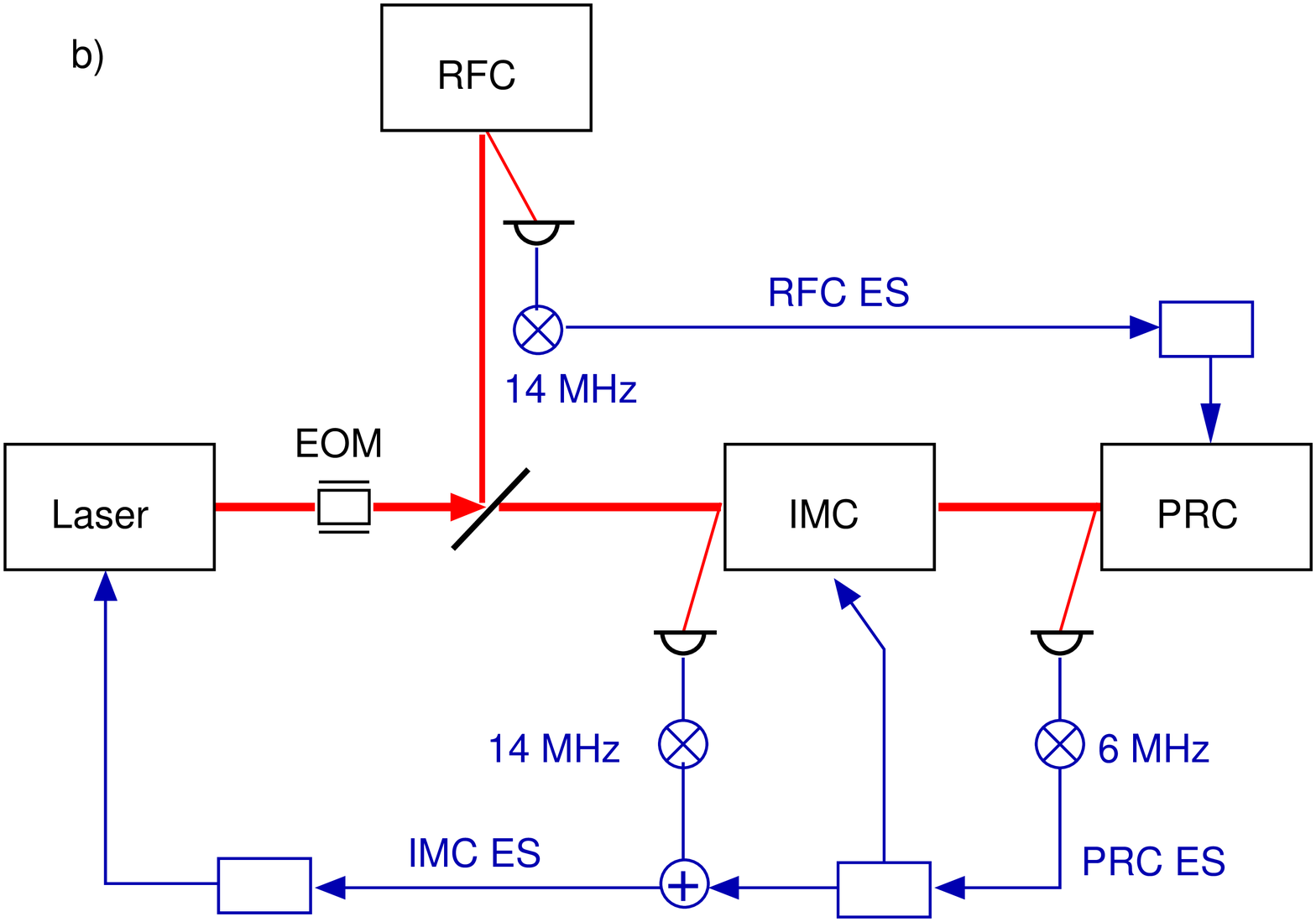}
\end{center}
\caption{\label{fig:double-loop}The control system for the laser frequency stabilisation: The
left schematic (a) shows the control for the pre-stabilisation, which is used during the
lock-acquisition phase, whereas on the right (b) the control in the final
configuration is depicted.  The laser light is modulated in phase 
at 6 and 14\,MHz before it enters the IMC, and the error signals
are derived using the Pound-Drever-Hall technique.
(a): In case of the pre-stabilisation, the light reflected by IMC and RFC is
detected and the signals are
demodulated at 14\,MHz. Properly filtered,
these signals are fed back to the laser and also to the length of the
IMC. Thus, the laser light is stabilised to the length of the IMC,
which follows the RFC length for low frequencies.
(b): After the lock of the main interferometer has been acquired the 
light reflected by the PRC is detected, demodulated at 6\,MHz. This
signal is then used to control the laser frequency and
the IMC length. The low-frequency stabilisation is achieved as before
except that the RFC signal is now fed back to the length of the PRC.}
\end{figure}
 
The laser source is an injection-locked master-slave system that can deliver 20\,W
of continuous power at 1064\,nm. The laser light is modulated in phase at a frequency of 14\,MHz before it
enters the vacuum.
The laser frequency is then stabilised to the length of the IMC up to a Fourier frequency of 270\,kHz
using a standard Pound-Drever-Hall (PDH) scheme. 
The bandwidth
of this control loop is limited by the FSR of the IMC at $\approx 1$\,MHz. As the optical
components of the IMC are suspended in vacuum this control provides an
excellent stability at high frequencies. In order to reduce the low-frequency 
fluctuations, 0.1\% of the light impinging on the IMC is split off and injected into
the RFC. Another PDH loop is used to control the length of the IMC with respect to 
the length of the RFC, for frequencies below 15\,Hz. Thus the laser frequency 
inherits the good DC stability of the RFC.
The pre-stabilisation was successfully implemented and tested during the commissioning
of the CITF~\cite{CITF}. 
\begin{figure}[h]
\begin{center}
\IG [scale=.5, angle=0] {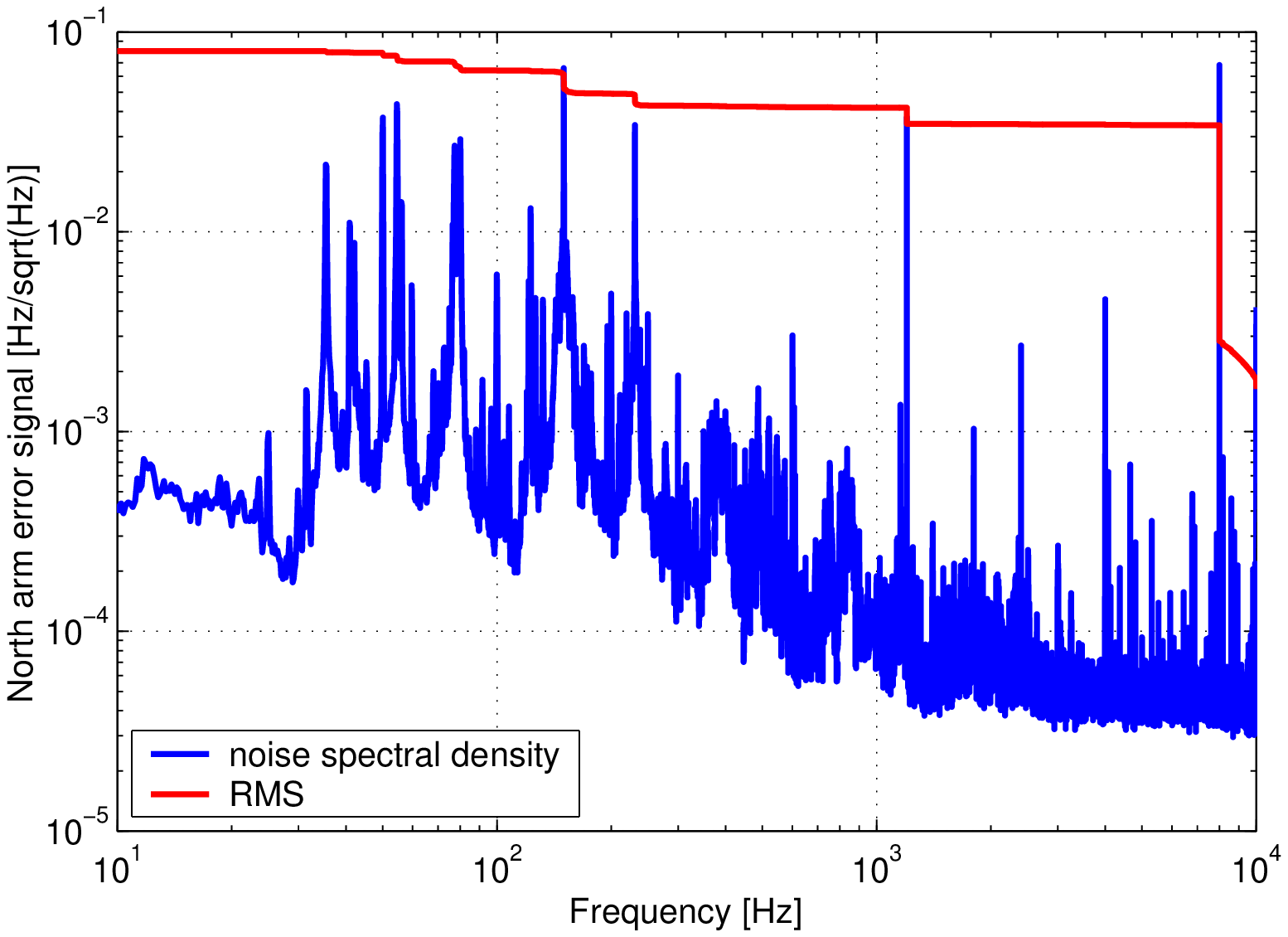}\hspace{4mm}\IG [scale=.5, angle=0] {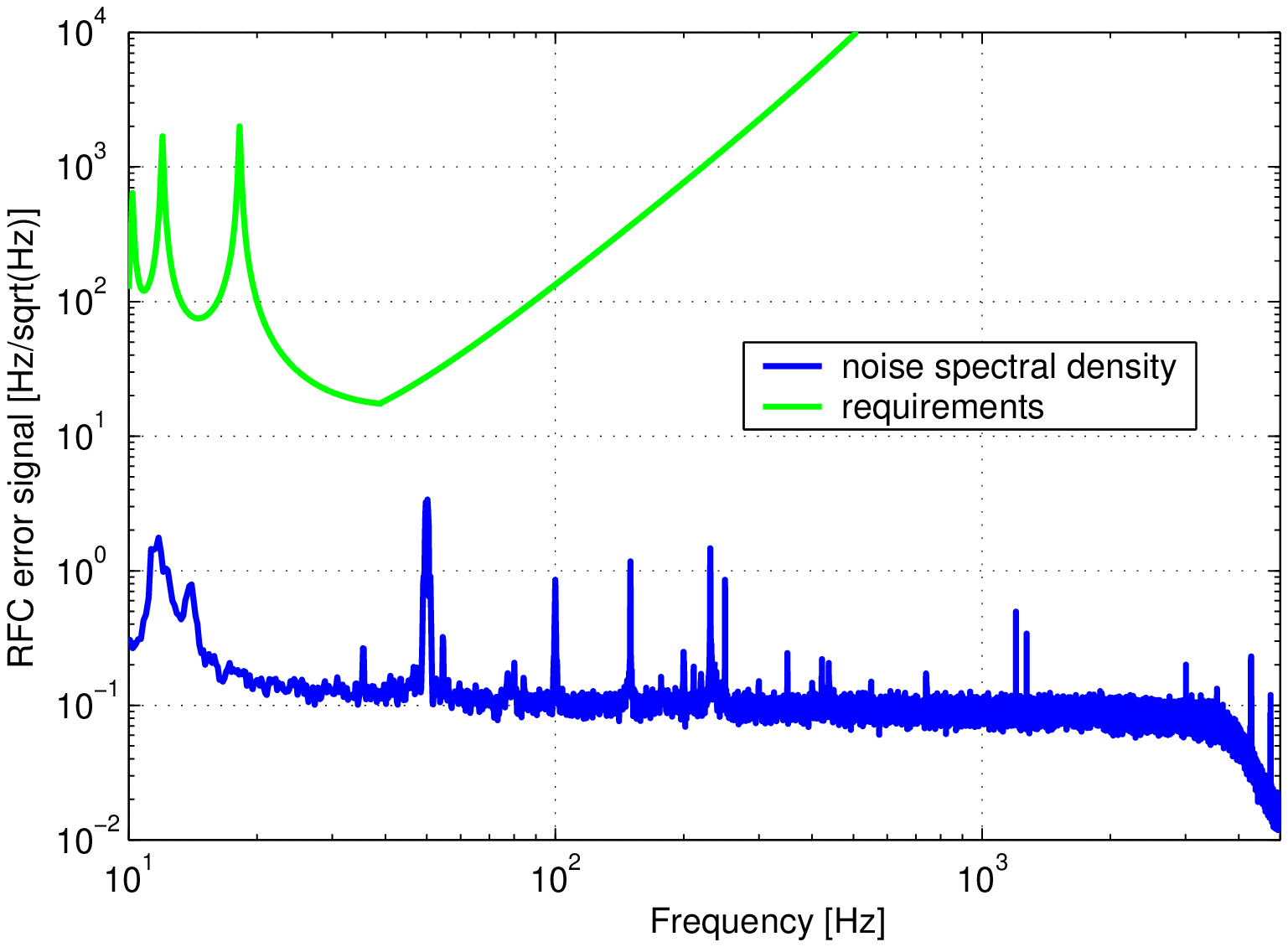}
\end{center}
\caption{\label{fig:freqnoise}The graphs show the error signals for the north arm (NA ES) on the left and for
the RFC (RFC ES) on the right during the C3 run: The north arm was used to test the SSFS.
The NA ES gives an RMS of less than 0.1\,Hz; the noise spectrum is still higher than the
requirements ($10^{-5}$\,Hz$/\sqrt{\rm Hz}$, see text), 
while the error signal for the RFC is well below the requirements.} 
\end{figure}

The combination of two control loops as described in the previous paragraph 
performs the laser pre-stabilisation 
needed during the lock acquisition of the main interferometer. As soon as the interferometer
is locked in all longitudinal degrees of freedom the laser frequency will be further
controlled to be resonant inside the \prc. 

The layout of the control system 
is shown in \mFig{fig:double-loop}. A second phase modulation at $\approx6$\,MHz
(the frequency is tuned to be resonant in the IMC) is applied to the laser light.
The laser frequency is first stabilised to the IMC as before.
The PDH error signal obtained with the light reflected from the \pom\ is added into the
error point of the IMC-loop (bandwidth 1\,kHz). In addition, the same signal is
applied to the length of the IMC (bandwidth 200\,Hz). This guarantees that the laser frequency and the
length of the IMC follow the length of the recycling cavity, which, due to
its greater length, provides a better relative stability for Fourier frequencies above the
internal resonances of the mirror suspensions. 
The incorporation of this error signal is called \emph{second stage of frequency stabilisation} (SSFS). 
The control signal derived from the reference cavity can no longer be 
applied to the IMC and is fed back to the length of the PRC.

The SSFS was implemented in a preliminary configuration in early 2004. 
Instead of the PRC, which is not yet available, the north
arm cavity (NA) was used. The NA has a simpler optical transfer
function than the PRC, and, in addition, the
injected power in this setup is 
lower than with recycling (due to the attenuation of the misaligned PR).
Nevertheless, it can be used to test the performance of the designed control system.
The first four days of the C3 run were dedicated to test the SSFS:  
The west arm cavity was not used and the laser was locked to the NA
using diode B1 (see \mFig{fig:optical-layout}) for the NA error signal (NA ES).
This signal was then used exactly as shown for PRC ES in \mFig{fig:double-loop}
and the low-frequency control signal from the RFC was fed back to the NE mirror.

The system worked reliably during the run, with the longest continuous operation
of 32 hours. \mFig{fig:freqnoise} shows the in-loop frequency noise (i.e. the
error signals of the RFC and of the NA) during the run. The RFC loop is already implemented
in its final configuration and the noise is well below the requirements.
The NA error signal is higher than the requirements 
(approximately $10^{-5}$\,Hz$/\sqrt{\rm Hz}$) because the lock of the laser to the NA uses only a 
simple filter. The final control loop will have more gain, and the
current measurements expect us to achieve the required frequency stabilisation. 

\section{Recombined Interferometer}
In the recombined configuration both arm cavities are longitudinally locked at resonance,
the \Mi\ is 
held on the dark fringe, and the PR mirror is strongly misaligned.
This mode can be used for various states of the other
subsystems, i.e. the frequency stabilisation, the automatic alignment and
the OMC. Although the MI in the recombined configuration does not provide
a very sensitive detector, it is a final step on the way to the
detector with recycling. It allows to characterise the optical signals
so that the proper control system for the recycled interferometer
can be derived.

Three longitudinal degrees of freedom have to be controlled.
Several methods are available to derive the necessary control signals from 
the various available photo diodes (see \mFig{fig:optical-layout}). 
By demodulating the photo current of B1, B1p, B2, B5, B7 and B8
at 6\,MHz one can extract error signals that are proportional 
to one or a combination of longitudinal degrees of freedom.
In addition to the suitable control signals a proper sequence for
the lock acquisition has to be found.

During the C3 run the following control scheme was used:
first the two arm cavities are independently locked to the
pre-stabilised laser. The error signal for the NA is obtained
from B5, and the error signal for the WA from B8. Both signals are
linearised by dividing them by the respective cavity power (detected on B7
and B8 respectively). This method extends the linear range
of the error signal by a factor of 10. The correction signal is fed back to 
the respective end mirrors. The MI dark fringe lock is switched
on; the error signal is generated by demodulating the signal from the dark fringe (B1)
at 6\,MHz; it also contains
the gravitational wave signal. In this case the OMC was not locked but
held on or close to a resonance by hand. The correction signal is applied to the
beam splitter. All three loops use the same filter with a bandwidth 
of 50\,Hz. The locking accuracy (of the MI control) that can be
achieved with the current control scheme has been estimated from
the measured data to  be $6\cdot10^{-12}$\,m$_{\rm RMS}$,
assuming that the SSFS will improve the frequency stability as expected.
This accuracy is already 
close to the requirements for the final setup, in which the accuracy will be improved 
considerably by the high optical gain.
\begin{figure}[t]
\begin{center}
\IG [scale=0.3, angle=0]{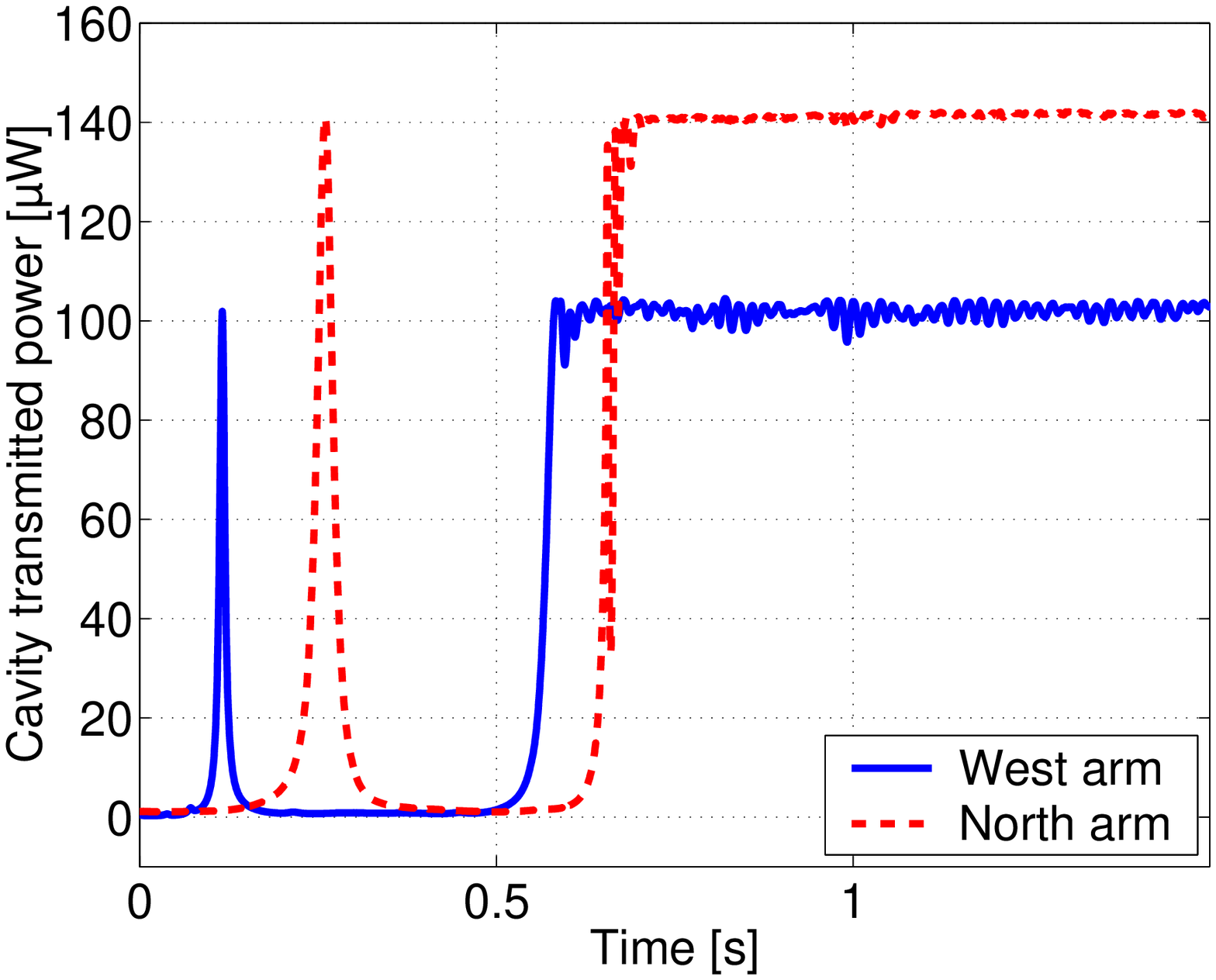}\hspace{3mm}\IG[scale=0.3, angle=0]{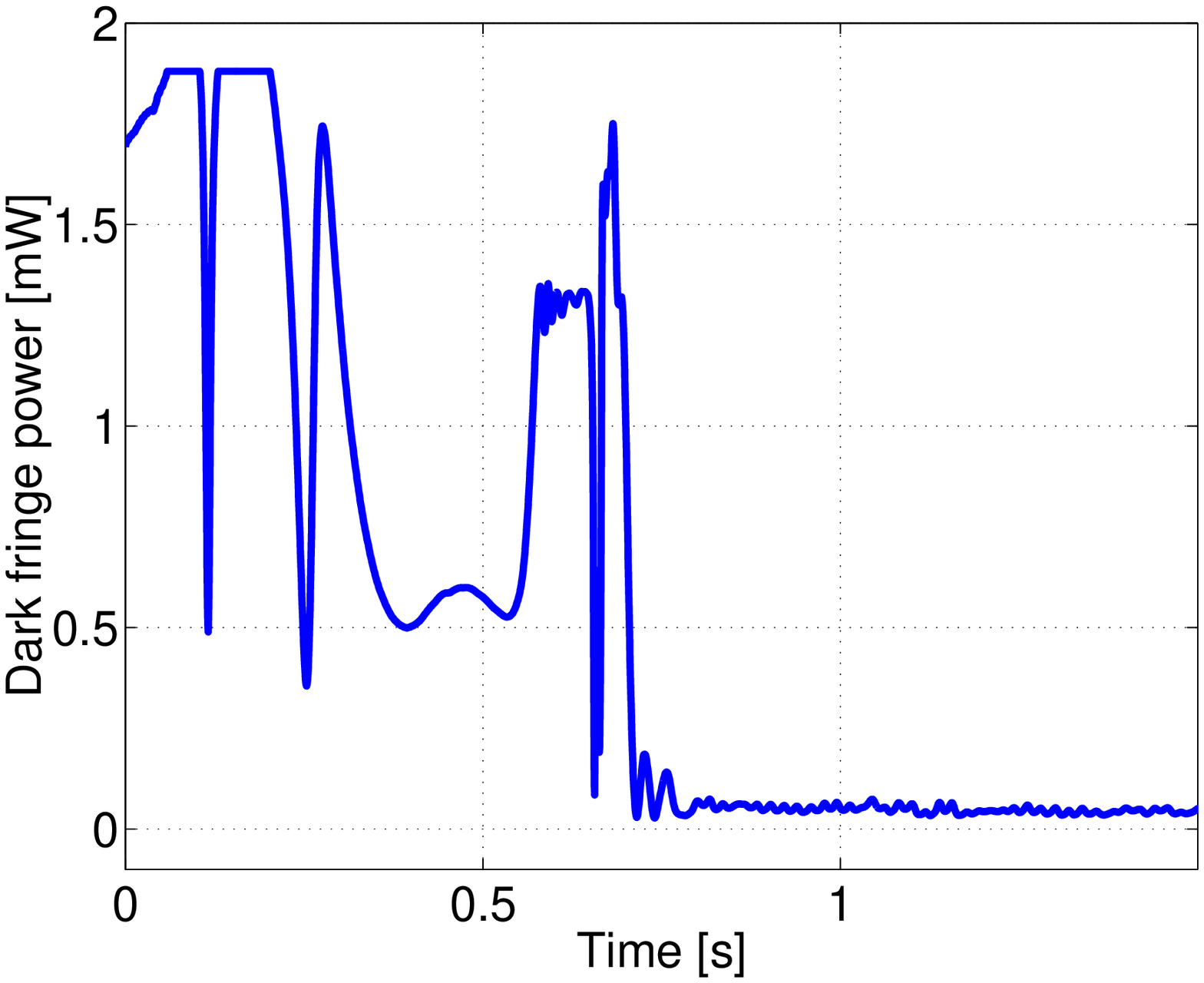}\hspace{3mm}\IG[scale=0.3, angle=0] {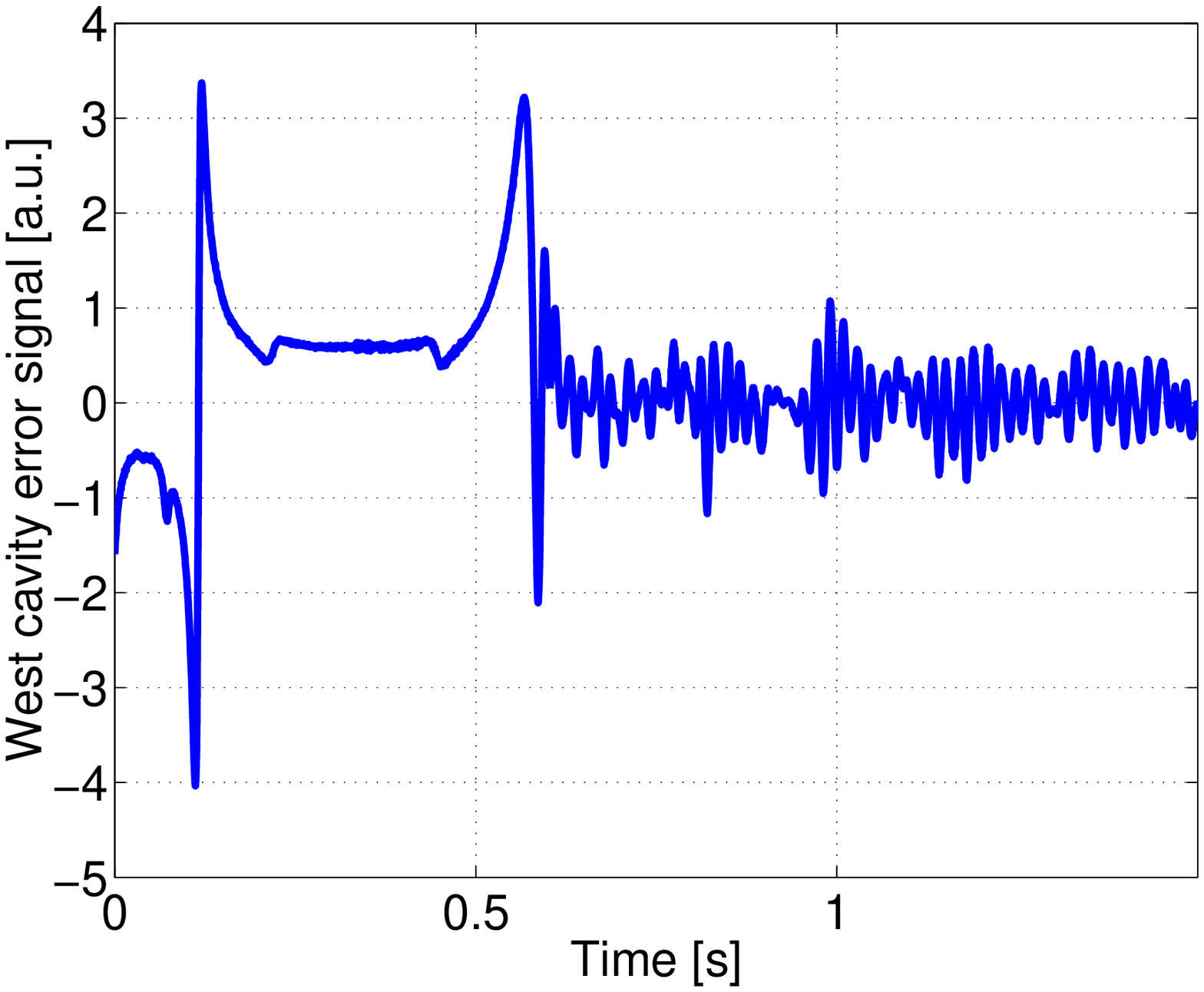}
\end{center}
\caption{\label{fig:recomb}A lock acquisition of the recombined interferometer:
The left plot shows the light power transmitted by the two arm cavities; the center
plot shows the dark fringe power and the right plot depicts the
error signal for the longitudinal control of the west arm cavity. At the
beginning of the data taking, the longitudinal control was off and both cavities passed
freely through resonances. At about 0.4\,s the control was switched on.
On the resonance both cavities are locked independently. This triggers
the control of the beam splitter so that a few tens of seconds after
engaging the control the MI is locked to the dark fringe.}
\end{figure}

The lock acquisition using this sequence is very reliable. \mFig{fig:recomb}
shows a time stretch of a lock acquisition event. Statistic
tests have shown that a successful acquisition of lock could be achieved
on every fringe. During the C3 run the recombined interferometer could
be locked continuously for several hours, limited by drifts
of the mirror alignment because the WA automatic alignment was not
yet implemented. The error signal for the MI control 
suffers from offsets which are introduced by misaligned optics.
Therefore, the automatic alignment must be implemented to
achieve a stable long-term operation of the recombined interferometer.

During C3 a different loop filter for the arm cavity control was also tested. This
filter has a steep roll-off at 300\,Hz in order to avoid re-injecting 
electronic noise at higher frequencies. The fourth plot in \mFig{fig:Csens} shows the
corresponding improvement in sensitivity when this filter is used.

\section{Automatic Alignment}
The optics of the \Mi\ and of the cavities have to be well-aligned
with respect to each other and with respect to the incoming beam 
to reach their high optical qualities.

To guarantee a stable long-term operation and a high sensitivity the angular
degrees of freedom have to be actively controlled. A standard technique 
for measuring the misalignment of optical components in this type
of interferometer is the so-called \emph{differential wave-front sensing}. 
It can be used with the light
reflected or transmitted by the cavity; the latter, also called
\emph{Anderson technique}~\cite{anderson}, is used in VIRGO~\cite{LA}. The laser light 
is modulated in phase at a specific RF 
frequency, such that the
sidebands in the first order transverse modes (\M{01}) are resonant
in the arm cavities. The transmitted light is detected by split photo detectors,
\ie quadrant diodes (QD) with four independent segments.
The photo currents of these QDs are demodulated at the RF frequency to
yield signals proportional to the misalignment of the optical components.
The two horizontal degrees of freedom (or
vertical respectively) can be separated by using two QDs
that are positioned to detect the light field at a different Gouy phase.  
The correction signals are fed back to the coils of the marionette. This,
and the low bandwidth of the control loop ($<5$\,Hz), ensures that the
angular control does not introduce noise in the measurement band.

\begin{figure}[h]
\begin{center}
\IG [scale=0.5, angle=0] {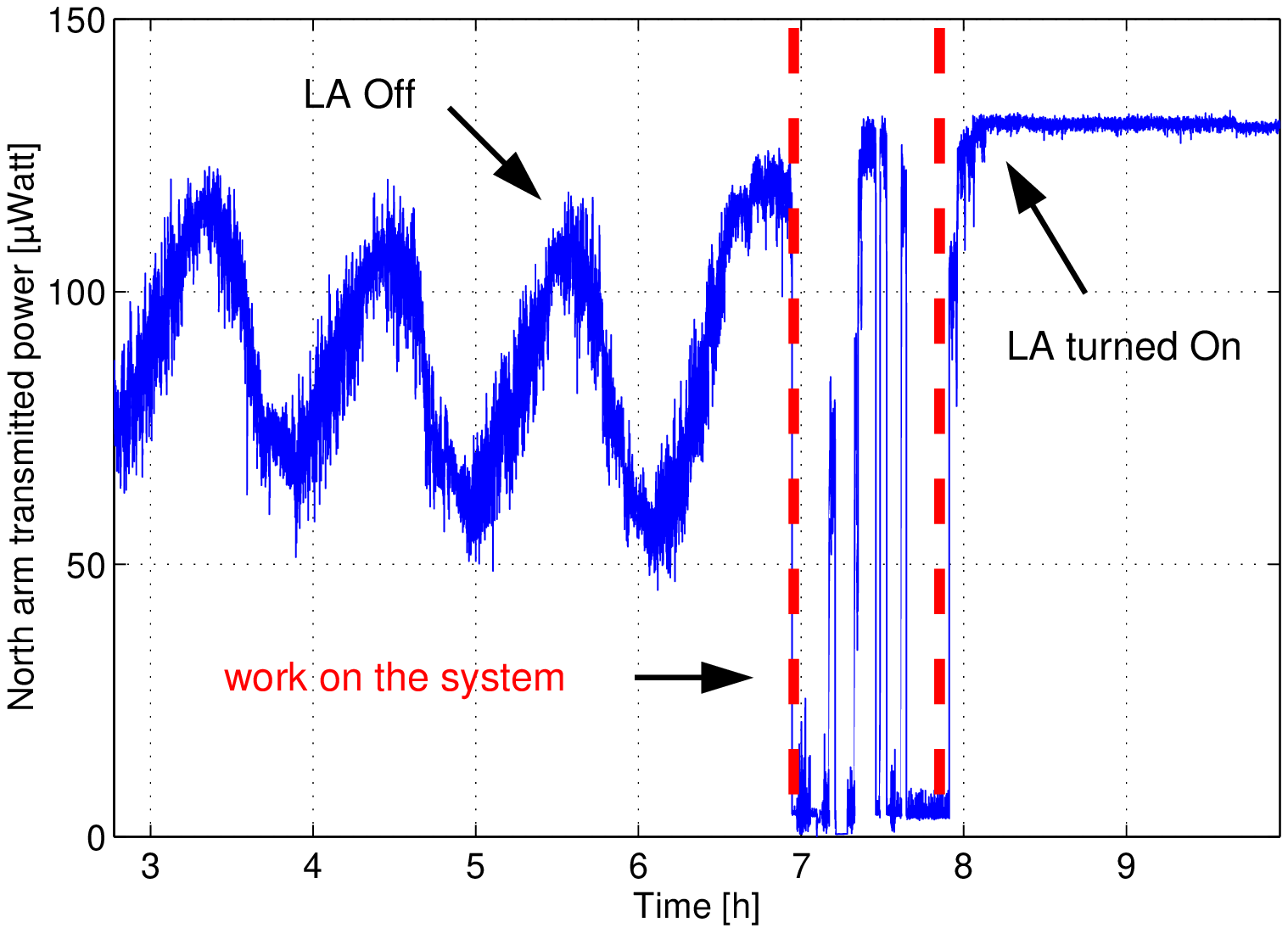}\hspace{7mm}\IG [scale=0.5, angle=0] {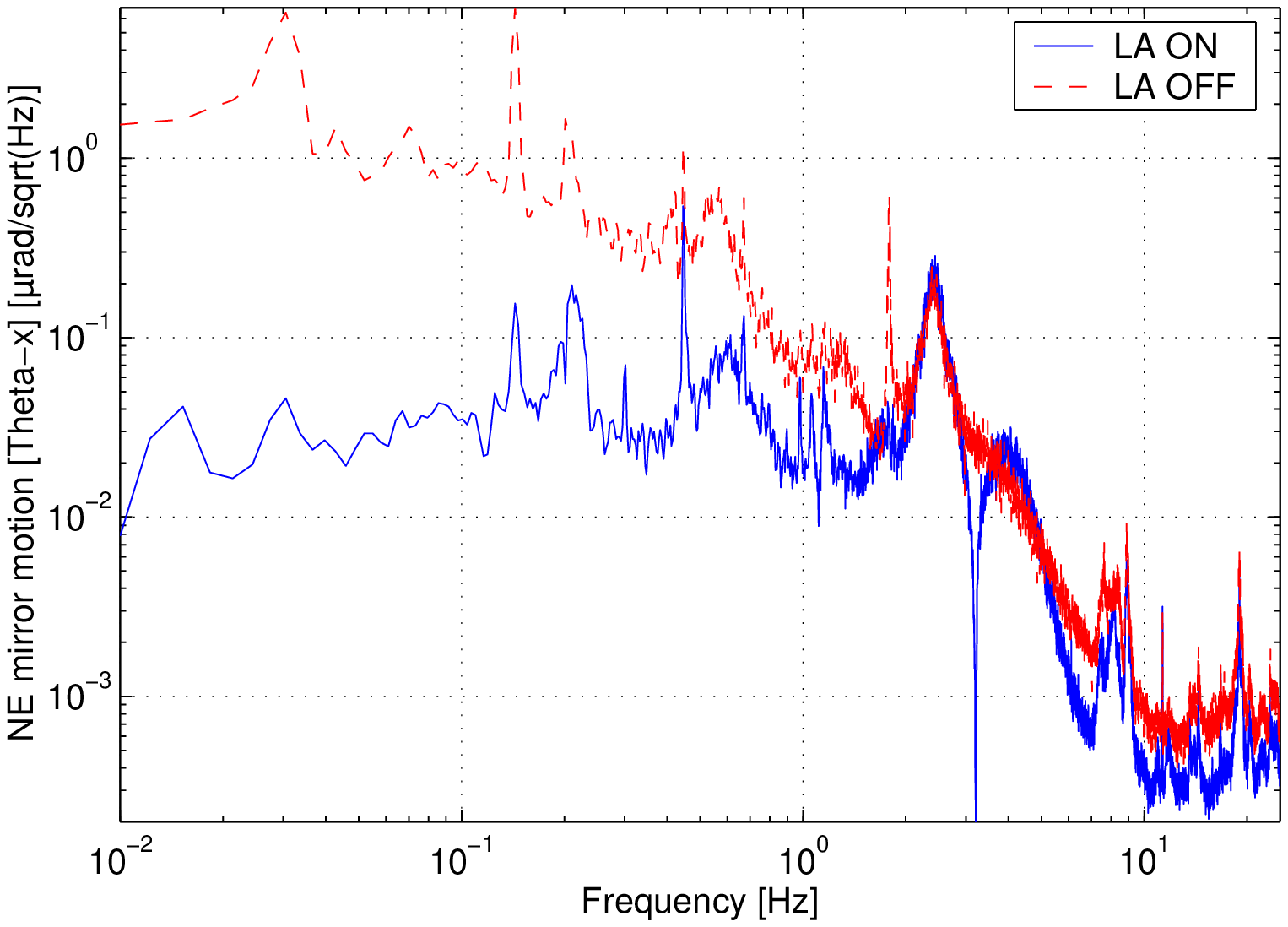}
\end{center}
\caption{\label{fig:PrB7Dc}Typical
performance of the automatic alignment control. The left plot shows the light power transmitted by the
north cavity as a function of time. On the left (time<7\,h) the automatic alignment was
switched off, then after a short period in which the system was prepared
for the C2 run, the cavity is controlled again, this time with the automatic 
alignment turned on so that the power fluctuations are reduced. The right plot compares 
the noise spectral density of the NE mirror motion in $\Theta_x$ for automatic alignment
switched on and off (measured in-loop). 
Below the unity gain frequency (around 3\,Hz)
the mirror motion is dominated by suspension resonances of the cavity mirrors or the
IMC. For higher frequencies the amplitude falls rapidly to $\approx1$\,nrad$/\sqrt{\rm Hz}$ at 10\,Hz. 
The noise floor for frequencies above 1\,Hz is correlated to the laser frequency noise.}
\end{figure}

During the C2 run (see \mFig{fig:PrB7Dc}) the automatic alignment of the north
arm cavity was tested with an otherwise unchanged setup. 
Thus the Anderson 
technique was demonstrated for the first time 
in a large-scale interferometer.
The presence of the alignment control reduces the power fluctuations in the cavity 
considerably and allows long continuous operation without
manual re-alignment of the optics. 

During C3 the automatic alignment of the
NA was tested together with the SSFS, and a 32-hour long period of continuous operation
shows the reliability also of the alignment control. Recently the automatic
alignment was successfully implemented on the west arm. This step completes
the automatic alignment for the current interferometer configuration. 

\section{Output Mode Cleaner}
Due to several possible optical defects, like mirror surface deformations, misalignments
and radii of curvature mismatch, the interference at the main beam splitter is
degraded; as a consequence, a fraction of the light leaves the
dark port as higher order TEM modes. This increases the shot-noise level 
on the main photo detector, and decreases the sensitivity. 
This problem can be solved with another cavity in front of the 
photo detector filtering out the higher order modes. This
\emph{\omc} (OMC) is a short monolithic cavity ($l=3.6$\,cm) made of silica.
It is located on the detection bench (DB) (see \mFig{fig:optical-layout}),
and has a finesse of 50 and a large bandwidth of 
75\,MHz. The cavity is controlled by changing
its length via the temperature using a Peltier cell that is
connected to the support structure. The error signal
is generated by modulating the cavity length
at 28\,kHz with a piezoelectric actuator. 
It is difficult for the
length control of the OMC to identify the correct mode, \ie the \M{00}, 
since a large amount of light might be in higher modes and the error signal
is similar for the resonances of all TEM modes. For this purpose
a CCD read-out system is used for analysing the transmitted light
of the OMC, and the transmitted mode is identified with a $\chi^2$ method.

The control and the automatic acquisition of lock of the OMC were already
demonstrated successfully during the CITF commissioning; however, in that
case the DB was fixed and not in vacuum. During the VIRGO commissioning
the OMC control and alignment were updated to work with the DB suspended in vacuum. 
Due to the vacuum the
control via temperature is slower; a lock of the OMC can take 10 to 20 minutes.
An automatic alignment system uses two quadrant diodes to automatically
center the beam coming from the beam splitter to the OMC by moving
two mirrors of a mode-matching telescope in front of the OMC. This method
is used when the
automatic alignment of the interferometer is engaged, because
otherwise the asymmetric shape of the beam impinging on the quadrants 
creates a false signal.

The OMC was used regularly during the commissioning runs. The lock, once acquired, 
is very robust. We did not observe any unlock due to the OMC itself.
When the main interferometer unlocks (and the IMC stays locked) 
the OMC temperature is stabilised to the current value so that the OMC lock can be 
recovered immediately after a re-lock of the interferometer.

The precision of the OMC length control has been measured to 
be $\lambda/60000$, which is 10 times better than the requirements. 
During the CITF commissioning the contrast improvement 
due to the OMC was measured  to be of the order of 10.


\section{Conclusion}
Since the start of the commissioning of the VIRGO detector in
September 2003, both 3\,km long arm cavities have been put
into operation. The longitudinal control, and the
automatic alignment have been implemented.
In February of this year the \Mi\ with the arm cavities
has been locked in the recombined configuration.

In addition, the second stage of frequency stabilisation has been
implemented and tested, using the north arm cavity. 
The detection bench with the \omc\ was suspended and put
in vacuum.

Three commissioning runs, periods of continuous data taking, 
have been performed to understand the performance of the system. 
Several upgrades and optimisations 
of the subsystems, mainly the active control of the IMC, have been
carried out to improve the sensitivity and reliability of the detector.

In June
the next commissioning run will be used to operate the
interferometer in the recombined configuration with automatic
alignment in all degrees of freedom and the second stage
of frequency stabilisation. 
This completes the last intermediate phase before moving
to the final configuration, including power recycling.
By the end of this year the detector should be fully 
operationable in this configuration.

\doitemsep

\end{document}